\documentclass[12pt]{iopart}
 \usepackage{iopams}  
 \usepackage{epsfig}

 \def \be{\begin{displaymath}}
 \def \ee{\end{displaymath}}              

 \def \ben{ \begin{equation} }
 \def \een{ \end{equation}   }            

 \def \bea{\begin{eqnarray*}}             
 \def \eea{\end{eqnarray*}}

 \def \bean{\begin{eqnarray}}             
 \def \eean{\end{eqnarray}}

 \def \nn{\nonumber}

 \def \Ref#1{(\ref{#1})}
 \def \fie {\varphi}
 \def \eps{\varepsilon}

 \def \inv{ ^{-1} }

 \def \invb#1 { \frac{1}{#1} }

 \def \fr#1#2{ \frac{#1}{#2} }

 \def \lapprox{ \stackrel{<}{\sim} }

\begin{document}

 \title[Coulomb drag between ballistic one-dimensional electron
   systems]{Coulomb drag between ballistic one-dimensional electron systems}

 \author{P. Debray${}^1$, V. Gurevich${}^2$, R.
   Klesse${}^3$, R. S. Newrock${}^4$}


 \address{${}^1$ Service de Physique de l\'Etat Condens\'e, CEA Saclay, 91191
 Gir-sur-Yvette, France}
 \address{${}^2$ Solid State Physics Division, A.~F.~Ioffe Institute,
   194021 Saint Petersburg, Russia}
 \address{${}^3$ Institut f{\"u}r Theoretische Physik, Universit{\"a}t
   zu K{\"o}ln, Germany}
 \address{${}^4$ Department of Physics, University of Cincinatti, Ohio
 45221-0011,
   USA}

\begin{abstract} 
 The presence of pronounced electronic correlations in a
 one-dimensional systems strongly enhances Coulomb coupling and is expected to
 result in
 distinctive features in the Coulomb drag between them that are absent
 in the drag between two-dimensional systems. In this article, we
 review recent Fermi and Luttinger liquid theories of Coulomb drag
 between ballistic one-dimensional electron systems, alias quantum
 wires, in the absence of inter-wire tunneling, to focus on these
 features and give a brief summary of the experimental work reported so
 far on one-dimensional drag. Both the Fermi liquid (FL) and the
 Luttinger liquid (LL) theory predict a maximum of the drag resistance
 $R_D$ when the one-dimensional subbands of the two quantum wires are
 aligned and the Fermi wave vector $k_F$ is small, and also an exponential
 decay of $R_D$ with increasing inter-wire separation, both features
 confirmed by experimental observations. A crucial difference between
 the two theoretical models emerges in the temperature dependence of
 the drag effect. Whereas the FL theory predicts a linear temperature
 dependence, the LL theory promises a rich and varied dependence on
 temperature depending on the relative magnitudes of the energy and
 length scales of the systems. At very low temperatures, the drag
 resistance may diverge due to the formation of locked charge density
 waves. At higher temperatures, it should show a power-law dependence
 on temperature, $R_D \propto T^x$, experimentally confirmed in a narrow
 temperature range, where $x$ is determined by the Luttinger liquid
 parameters. The spin degree of freedom plays an important role in the
 LL theory in predicting the features of the drag effect and is
 crucial for the interpretation of experimental results. Substantial  
 experimental and theoretical work remains to be done for a
 comprehensive understanding of one-dimensional Coulomb drag.
 \end{abstract} 



 \maketitle

\section{Introduction}

 Moving charge carriers in a conductor exert a Coulomb force on the charge 
 carriers in a nearby conductor and induce a drag current in the latter via 
 momentum transfer. This phenomenon, known as Coulomb drag, was predicted by 
 Pogrebinskii in his pioneering paper \cite{progrebinskii} in which he
 argued that in a  structure of two semiconductor layers separated by
 an insulating layer, there would be a drag of carriers in layer 1
 ("drag layer"), resulting in a   
 drag current $I_{D}$,\textit{} due to the direct Coulomb interaction
 with the carriers
 in layer 2 ("drive layer"), where an electric current $I$ flows. If no current 
 is allowed to flow in the drag layer, the charge carriers will accumulate at 
 one end inducing a charge imbalance across the layer. This charge will 
 continue to accumulate until the force of the resulting electric field 
 balances the drag force. In the stationary state there will be an induced or 
 `drag' voltage $V_{D}$ in the drag layer. When the carriers in both
 layers are of the same type (electrons or holes), the drag voltage has
 a sign opposite to the voltage drop in the drive layer.
 Figure \ref{fig-schematic}
 gives a schematic view of  
 Coulomb drag between two parallel quantum wires. The quantity usually 
 measured in experiments is the drag voltage $V_{D}$. The drag resistance 
 $R_{D}$ is defined as $R_{D}$ = -$V_{D}$ /$I$. 
  \begin{figure}[htb] 
  \begin{center}
          \epsfxsize 5cm
  \end{center}
  \caption{Schematic view of Coulomb drag between parallel quantum wires.}
  \label{fig-schematic}
  \end{figure}

 Coulomb drag between two-dimensional (2D) electron systems has been 
 extensively studied \cite{rojo} both experimentally and theoretically. The basic 
 physics involved in the description of the drag in two dimensions is now well 
 understood on the basis of Fermi liquid (FL) theory of interacting fermions. 
 The FL theory is well established in three dimensions, holds
 marginally for many two-dimensional systems, but generally fails in
 one-dimension. 
 The theory is based on Landau's
 conjecture that the low-lying  excitations of interacting fermion
 systems can be connected continuously to those of the non-interacting
 Fermi gas -- there is a smooth mapping between the quasiparticles of
 the interacting and of the non-interacting system \cite{nozieres}.

 Coulomb drag between one-dimensional (1D) electron systems has been the 
 focus of considerable interest in recent years because our understanding of 
 the quantum properties in interacting 1D systems is unsatisfactory. 
 Experimental work on the subject remains quite limited
 \cite{moon,debray2000,debray2001,yamamoto}, a fair number of 
 theoretical papers have been published
 \cite{GPF,sirenko,hu_flensberg,tanatar,komnik,flensberg,RV,nazarov,ponomarenko,klesse_stern,GM}.
  The primary reason for this 
 theoretical interest is that Coulomb drag is one of the most effective ways 
 to study electron-electron (e-e) interaction.

  It is now theoretically 
 established that in an interacting 1D electron gas of infinite length
 the e-e-interaction completely  
 modifies the ground state of the system. The elementary excitations can not 
 be treated as non-interacting quasiparticles of a conventional Fermi liquid, 
 but instead acquire a bosonic nature. An adequate theoretical description of
 these interacting 1D systems can be done in terms of the so-called
 Luttinger-liquid (LL) \cite{haldane} (for a recent review see
 \cite{voit}), complementary to the FL description in higher dimensions.
In a real 1D system of finite length at finite temperatures, the
extent of influence of the e-e interaction will depend on the system
parameters. 

 Experimental efforts  to observe manifestation of Luttinger liquid
 behavior, however, have been quite  
 limited \cite{tubes,auslaender,takiainen,auslaender02}.
 Part of the problem is that the (e-e) interaction has little influence on the 
 conductance of a single wire, since the current is proportional to the total 
 electron quasi-momentum, which is conserved in electron-electron collisions. 
 To look for experimental evidence of the LL state it would therefore
 help to explore a new
 experimental tool based on new devices and physical phenomena. Coulomb drag 
 between 1D electron systems in a dual-wire configuration 
 and opens up a new opportunity and avenue for experimentally probing
 the LL state in a 1D electron system.

 The purpose of this review paper is to present the current status of the 
 theory of Coulomb drag between 1D electron systems for electron transport in 
 the ballistic regime, and to report on experimental measurements of the 1D 
 drag effect. Ballistic transport takes place when the quantum wire 
 dimensions are smaller than both the elastic and the inelastic
 scattering lengths. 
 Electron transport is strictly one dimensional when only 
 the lowest 1D subband of the wire is occupied and transport takes place 
 in the fundamental mode. The ballistic regime is well suited for Coulomb 
 drag study, since in this regime other scattering processes, such as impurity 
 and phonon scattering, are either insignificant or totally absent.

 Further theoretical and experimental investigation of the 1D Coulomb drag 
 effect can enhance our general understanding of the properties of systems of 
 low dimensionality. This broad class of systems is currently a very active 
 area of research. In addition to its fundamental interest, a comprehensive 
 understanding of Coulomb interaction between quantum wires is expected to 
 play a significant role in the design of nanodevices, such as 
 single-electron transistors (SETs) \cite{tsukagoshi} and quantum
 cellular automata (QCA) \cite{amlani}, which are comprised of quantum
 dots and quantum wires in close proximity.

 The paper is organized as follows. Section \ref{sec-fermi} describes the theory
 of 1D 
 Coulomb drag based on the Fermi liquid approach. Section
 \ref{sec-luttinger} reviews the 
 Luttinger liquid description of this effect. In Section \ref{sec-exp} is
 presented a 
 summary of the experimental work reported so far and a comprehensive 
 analysis of the experimental results using both the FL and the LL 
 descriptions of 1D Coulomb drag. Finally, Section \ref{sec-future}
 gives some guidelines for future work on the subject.

\section{Fermi liquid approach}\label{sec-fermi}
 In this section the 1D Coulomb drag is analyzed within
 the Fermi-liquid concept. We will follow Refs.~\cite{GPF,GM} 
 and use the physical picture developed by Landauer~\cite{Lan},
 Imry~\cite{Imry}, and B\"uttiker~\cite{Buttiker}.
 We assume that each quantum wire is connected to ideal electronic reservoirs
 attached to its ends. The relaxation processes in the reservoirs are
 considered to be so fast that each of them is in thermal equilibrium.

 The e-e interaction within a single quantum wire does not result in a
 current variation because 
 of the quasimomentum conservation in the e-e collisions. 
 However, if two such wires, 1 and 2 are near one another and
 are parallel, the Coulomb interaction of electrons belonging to
 different wires can transfer momenta between the wires, 
 which eventually gives rise to a drag effect.

 The drag force due to
 the ballistic current in wire 2 creates a sort of permanent
 acceleration on the electrons of wire 1. As wire 1 has a
 finite length $L$ a steady drag current $J$ is established.

 Within the Fermi liquid approach we restrict ourselves to 
 direct electron-electron collisions mediated by the Coulomb interaction.
 Let us analyze the conservation laws for such collisions of electrons
 belonging to two different wires, 1 and 2, each of them being parallel
 to the $x$-axis. 
 We have
 \begin{equation}
 \epsilon^{(1)}_{nk}+\epsilon^{(2)}_{n'k'}=
 \epsilon^{(1)}_{l,k+q}+\epsilon^{(2)}_{l',k'-q}
 \label{1}
 \end{equation}
 Here $\hbar k$ is the $x$-component of the of the electron quasimomentum.
In this and the next sections we will put $\hbar=1$ and $k_{\rm B}=1$ (where
$k_{\rm B}$ is the Boltzmann constant). These quantities will be restored only in 
some final (or most important) formulas. Now, 
 \begin{equation}
 \epsilon^{(1,2)}_{nk}=\varepsilon^{(1,2)}_{n}(0)+k^2/2m ,
 \label{1a}
 \end{equation}
 $m$ being the effective mass while $n$ being the transverse
 quantization subband (channel) index, with primed quantities
 corresponding to wire 2 throughout.
 The solution of Eq. (\ref{1}) can be written as
 \begin{equation}
 q=-(k-k')/2\pm\sqrt{(k-k')^2/4+m\delta\varepsilon}
 \label{4}
 \end{equation}
 with
 $\delta\varepsilon=\varepsilon^{(1)}_{n}(0)+\varepsilon^{(2)}_{n'}(0)-
 \varepsilon^{(1)}_{l}(0)-\varepsilon^{(2)}_{l'}(0)$.

 We assume that the electrons of the quantum wires are
 degenerate and the temperature is low compared
 to the electron Fermi energy. For the electron-electron
 collision to be possible, the absolute values of the four
 quantities, namely, $\epsilon^{(1)}_{nk}, \epsilon^{(2)}_{n'k'},
 \epsilon^{(1)}_{l,k+q},$ and $\epsilon^{(2)}_{l',k'-q}$ should
 be within the stripes $k_{\rm B}T$ near the corresponding Fermi
 levels. This means that within the accuracy $mT/k_F$ the
 following relations should be valid
 \begin{eqnarray}
 k&=&k_F^{(n)},\quad k'=k_F^{(n')},\nonumber\\
 \vert k+q\vert&=&k_F^{(l)},\quad \vert k'-q\vert=k_F^{(l')} .
 \label{4''}
 \end{eqnarray}
 Here $k_F^{(n)}$ denotes the Fermi quasimomemtum for band $n.$ In
 general it is impossible by variation of a single quantity, i.e. the
 transferred quasimomentum $q$, to satisfy both relations of
 Eq. (\ref{4''}) (provided, of course, that the distances between the
 channel bottoms are much bigger than $T.$)

 In other words, one cannot in general satisfy Eq. (\ref{1}) for a
 finite $\delta\epsilon$. Therefore for a general case one should have
 $n=l ,\quad n'=l' .$ If both wires are identical equations $n=l'
 ,\quad n'=l$ are also possible. In both cases $\delta\epsilon=0$. We
 will assume the wires to be different.  Then
 \begin{equation}
 \delta(\epsilon^{(1)}_{nk}+\epsilon^{(2)}_{n'k'}-
 \epsilon^{(1)}_{l,k+q}-\epsilon^{(2)}_{l',k'-q})=
 (m/\vert q\vert)\delta(k-k'+q).
 \label{6'}
 \end{equation}
 This means that the quasimomentum transferred during a collision is
 $q=k'-k$, i.e. the electrons swap their quasimomenta as a
 result of collision.

 Assuming that the drag current in wire 1 is much smaller than the
 ballistic current in wire 2, we calculate the drag current by solving
 the Boltzmann equation for wire 1 (otherwise we should have solved a
 system of coupled equations for both wires). We assume the wires to
 be different though having the same lengths $L$ and consider the
 interaction processes when electrons in the two quantum wires after
 scattering remain within the initial subbands
 $\epsilon^{(1)}_{nk}=\varepsilon^{(1)}_n(0)+k^2/2m$ and
 $\epsilon^{(2)}_{n^{\prime}k}=\varepsilon^{(2)}_{n^{\prime}}(0)+k^2/2m$,
 $n$ being the subband's number. The Boltzmann equation for the
 electrons occupying the $n$th subband is
 \begin{equation}
 v_{k}\frac{\partial F^{(1)}}{\partial k}=
 {\cal I}^{(12)}\{  F^{(1)}, F^{(2)}\}
 \label{B1}
 \end{equation}
 where $F^{(1,2)}$ are the electron distribution functions in
 wires 1 and 2 respectively, and $I$ is the collision integral. We
 assume that the only type of 
 collisions that is essential is the interwire {\em e-e}
 collisions described by the term
 \bean
{\cal I}^{(12)}& &\{F^{(1)}, F^{(2)}\} = 2\int {dk'\over2\pi}
 \int {dq\over2\pi} \nn \\
 & & \sum_{n'}w(1,k+q,n; 2,k'-q,n'\leftarrow 1,k,n;2,k',n'){\cal P}
 \label{B2}
 \eean
 where
 \bean\label{3a}
 {\cal P} &=& \left[F^{(1)}_{nk}F^{(2)}_{n^{\prime}k^{\prime}}
 \left(1-F^{(1)}_{nk+q}\right)\left(1-F^{(2)}_{n^{\prime}k^{\prime}-q}
 \right)\right. \nn \\ 
 & & -
 \left.F^{(1)}_{nk+q}F^{(2)}_{n^{\prime}k^{\prime}-q}
 \left(1-F^{(1)}_{nk}\right)\left(1-F^{(2)}_{n^{\prime}k^{\prime}}\right)
 \right].
 \eean
 2 is the spin factor; the scattering probabilities are assumed
 to be spin-independent. If the e-e collisions can be treated
 within the perturbation theory then the scattering probability is
 given by 
 \bean
  w &(& 1,k+q,n, 2,k'-q,n'\leftarrow 1,k,n;2,k',n') = \nn\\
 & & 2\pi
 \vert\langle1,k+q,n; 2,k'-q,n'\vert
 V\vert 1,k,n;2,k',n'\rangle\vert^2 
 \times \nn \\
 & &\delta(\epsilon^{(1)}_{nk}+\epsilon^{(2)}_{n'k'}-
 \epsilon^{(1)}_{n,k+q}-\epsilon^{(2)}_{n',k'-q}) .
 \label{B3}
 \eean
 The matrix element of electron-electron interaction can be
 transformed to
 $$
 \langle1,k+q,n; 2,k'-q,n'\vert
 V\vert 1,k,n;2,k',n'\rangle
 $$
 \bean
 &=&{1\over L}\int d^2r_{\bot}\int d^2r'_{\bot}
 \phi_n^*({\bf r}_{\bot})\phi_{n'}^*({\bf r'}_{\bot}) \times \nn \\
 & & \quad V_{q}({\bf r}_{\bot}-
 {\bf r'}_{\bot})
 \phi_n({\bf r}_{\bot})\phi_{n'}({\bf r'}_{\bot})
 \label{B4}
 \eean
 where
 $V_{q}=\int dx V(x, {\bf r}_{\bot})\exp(-iqx),\,
 {\bf r}_{\bot}=(y,z).$
 We have
 \begin{equation}
 \int \;dx \;\int \;dx' \;V({\bf r}-{\bf r'})e^{iq(x-x')}=2e^2L
 K_0(\vert q 
 \vert\,\vert\Delta r_{\bot}\vert)
 \label{B6}
 \end{equation}
 where $\Delta {\bf r}_{\bot}={\bf r}_{\bot}-{\bf r'}_{\bot}$ and
 $K_0$ is a modified Bessel function defined in Ref.~\cite{RG}.
 Now,
 \begin{equation}
 K_0(\xi)={\left\{{-\ln(\xi/2),\quad \xi\ll1,}
 \atop{\sqrt{\pi/2\xi}\,e^{-\xi}},\quad \xi\gg1.\right.}
 \label{B6'}
 \end{equation}
 It means that the {\it e-e} interaction goes down exponentially
 provided $\vert q\vert\vert{\bf r}_{\bot}-
 {\bf r'}_{\bot}\vert/\hbar\gg1.$

 To calculate the current in wire 1 we iterate the Boltzmann
 equation (\ref{B1}) in the collision term ${\cal I}^{(12)}$.  The first
 iteration gives for the nonequilibrium part of the distribution
 function $\Delta F_{np}^{(1)}$
 \begin{equation}
 \label{df_1}
 \Delta F^{(1)}_{nk}=-\left(z\pm {L\over{2}}\right){1\over{v_{n}}}
 {\cal I}^{(12)}\{F^{(1)},F^{(2)}\}
 \end{equation}
 for $k\,>\,0$ ($k\,<\,0$)
 respectively. One gets for the drag current
 \begin{equation}\label{current}
 I_D=-2eL\sum_{n}\int_0^{\infty}
 {dk\over2\pi}{\cal I}^{(12)}\{F^{(1)},F^{(2)}\}.
 \end{equation}

 We assume in the spirit of the Landauer-B\"uttiker-Imry approach the
 driving wire connected to the reservoirs which we call
 ``left" ($+$) and ``right" ($-$), each of these being in independent
 equilibrium. Let the $x$-component of the quasimomentum of an
 electron in wire 2 before scattering be $k'$, after 
 scattering by an electron of wire 1 be $k'-q$. Let $k'>0$ while
 $k'-q<0 .$ Then the first distribution function in wire 2 is
 $F^{(0)}_{n'k}=f(\epsilon^{(2)}_{n'k}-\mu^{(+)})$ where $f$ is the
 equilibrium Fermi function. The second one is
 $F^{(0)}_{l',k+q}=f(\epsilon^{(2)}_{l',k+q}-\mu^{(-)})$ where
 $\mu^{(\pm)}=\mu\pm eV/2.$ At $eV=0$ the wires are in equilibrium. We
 denote the corresponding equilibrium chemical potential as $\mu.$

 Let us denote by $\Delta\{F\}$ the expression one gets after
 substitution of the equilibrium distribution functions given
 above into the collision term. For $k'>0$ ($k'<0$) and $k'-q<0$
 ($k'-q>0$) where $k'$ is the electron quasimomentum in wire 2
 before the scattering we obtain
 $$
 \Delta\{F^{(1)}, F^{(2)}\}=\pm2\sinh\left(\frac{eV}{2T}\right)
 $$
 \bean
 &\times& [1-f(\epsilon^{(1)}_{nk}-\mu)]
 [1-f(\epsilon^{(2)}_{n',k'}-\mu^{(+)})] \nn\\
 &\times& f(\epsilon^{(1)}_{n,k+q}-\mu)
 f(\epsilon^{(2)}_{n',k'-q}-\mu^{(-)}).
 \label{9}
 \eean

 We begin with a discussion of the Ohmic case $eV/T\ll1$.
 Accordingly, we replace
 $\sinh(eV/2T)$ by its argument and all the chemical potentials in
 Eq. (\ref{9}) by the same value $\mu$. The initial and final states
 of the colliding electrons should be within 
 $T$ of the Fermi levels~\cite{GPF}. This means that
 only the terms with $\varepsilon^{(1)}_{n}(0) = \varepsilon^{(2)}_{n'}(0)$
 where the equality is satisfied with the indicated accuracy give the
 principal contribution to the current. (The importance of equal
 channel velocities was also pointed out in Ref.~\cite{sirenko}). The
 contribution of each such pair of levels to the current is
 \bean
 I_D &=& \frac{e^5m^3Lk_{\rm B}TeV}{2\pi^2\kappa^2}\frac{1}{{k_n}^3}
 g_{nn}(2k_n)\frac{[\varepsilon^{(1)}_{n}(0)-\varepsilon^{(2)}_{n'}(0)]^2}
 {4(k_{\rm B}T)^2} \nn \\
 & & \times
 \left[\sinh{\frac{\varepsilon^{(1)}_{n}(0)-\varepsilon^{(2)}_{n'}(0)} 
 {2k_{\rm B}T}}\right]^{-2}
 \label{19''}
 \eean
 where
 \ben
  g_{nn'}(q) =  \left | \int d^2 r_{\bot}\int d^2 r'_{\bot}
 \vert\phi_n(r_{\bot})\vert^2\vert\phi_n(r'_{\bot})\vert^2
 K_0(q\vert \Delta {\bf r}_{\bot}\vert) \right |^2,
 \label{15}
 \een
 $\kappa$ is the dielectric susceptibility of the lattice,
 $k_n=\sqrt{2m[\mu-\varepsilon_n(0)]}$. This equation has been also
 obtained using linear response theory. In the case
 considered, wire 2 is a part of a usual structure for measuring
 ballistic conductance, i.e.  it joins two classical reservoirs, each 
 of them being in independent equilibrium. The driving current
 is~\cite{Imry}
 \begin{equation}
 I ={\cal N}\frac{e^2}{\pi}V,
 \label{20}
 \end{equation}
 $\cal N$ being the number of active channels (i.e. the subbands whose
 bottoms are below the Fermi level). So far a simplifying assumption
 has been used: the chemical potential $\mu$ in wire 1 and the
 average chemical potential in wire 2 are equal. In the general case
 they can have different values $\mu^{(1)}$ and $\mu^{(2)}$,
 respectively. Then one still gets Eq. (\ref{19''}) with the
 replacement $$
 \varepsilon^{(1,2)}_{n}(0)\rightarrow\tilde{\varepsilon}^{(1,2)}_{n}(0)
 \equiv\varepsilon^{(1,2)}_{n}(0)- \mu^{(1,2)} .
 $$

 One can measure either the current or the voltage that builds up in wire
 1.  The ratio of the drag current to the ballistic driving current for
 $\tilde{\epsilon}^{(1)}_n(0)=\tilde{\epsilon}^{(2)}_{n'}(0)$ is given by
 \begin{equation}
 \frac{I}{I_D}=\frac{4e^4m^3Lk_{\rm B}T}{\pi\hbar^3\kappa^2{\cal N}}
 \sum_{nn'}{\cal D}_{nn'}
\label{d1}
 \end{equation}
 where
 \ben
 {\cal D}_{nn'}=\frac{1}{\left(k_n^{(1)}+k_{n}^{(2)}\right)^3}
 g_{nn'}\left(k_n^{(1)}+k_{n}^{(2)}\right) .
 \label{21}
 \een
 Here $k_n^{(1,2)}=\sqrt{2m[\mu^{(1,2)} - \varepsilon^{(1,2)}_n(0)]}$.
 In this approximation, $k_n^{(1,2)}=k_{n'}^{(1,2)}$.
 In an experiment one usually measures the drag resistance $R_D = -V_D/I
 = I_D G_D/I$, where $G_D$ is the ballistic resistance of the drag wire and
 depends on the number of occupied subbands.  
 
 The 1D subband structure of the wires can be modified by changing the
 effective wire widths by applying appropriate gate voltages
 (Fig. \ref{fig-5}). 
 The variation of gate voltage may affect
 the positions of the levels of transverse quantization in the two
 wires in a different way. In the course of such a variation a
 coincidence of a pair of such levels in the two wires may be reached.
 The estimate (\ref{21}) is not very sensitive to the form of
 confining potential and electron densities.  In Fig. \ref{fig-3} the ratio 
 ${I}/{I_D}$ is plotted (for $\mu^{(1)}=\mu^{(2)}$) as a function 
 of the ratio of effective wire widths.  This plot exhibits striking
 oscillations with large peak-to-valley ratios. The peaks occur
 when channel velocities in two interacting wires are equal which
 happens whenever any two current-carrying channels line up. This sort
 of coupling is particularly strong when such channel velocities are
 quite small.  

 The condition
 $\tilde{\epsilon}^{(1)}_n(0)=\tilde{\epsilon}^{(2)}_{n'}(0)$ gives the
 main maxima of the drag current, especially for the lowest levels of
 transverse quantization. Some subsidiary maxima can also be observed,
 particularly in external magnetic field --- see Ref.~\cite{RV}.

 So far we have considered the peaks of the drag current under conditions where 
the Fermi
level is well above the coinciding bands in wires 1 and 2. Now we would like to 
say a 
few words about a special case that may be particularly important regarding the 
experiment described below.                                                         
 This is the case where the bottoms of two subbands not only
coincide but also just touch the Fermi level. Then the conduction electrons obey the 
so-called {\em intermediate statistics.} It means that their equilibrium distribution 
functions are
\begin{equation}
f={1\over{\displaystyle \exp(k^2/2mT)}+1}
\label{F1}
\end{equation}

The drag current is proportional to the {e-e} scattering probability averaged over this 
distribution function. The scattering probability itself is determined by the quantum 
mechanics, i.e. it depends on the form of electron wave function in the quantum wire, in
other words, on the exact form of the confining potential. Investigation of the 
temperature dependence of the drag in this special case is one of the problems of the 
theory to be solved in future.

 Now we turn to the case of non-Ohmic transport in the drive
 wire, i.e. to the case where $eV\gg k_{\rm B}T$~\cite{GM}.  The
 situation for $\mu^{(1)}=\mu^{(2)}$ is illustrated in
 Fig. \ref{fig-2}. The upper and the lower dashed
 lines correspond to the positions of the chemical potentials
 $\mu^{(-)}$ and $\mu^{(+)}$, respectively, while the middle dashed
 line corresponds to the 
 average value $\mu$. Parabolas (1) and (2) represent the dispersion
 law of electrons in wires 1 and 2 respectively.  The full circles
 correspond to the initial states of colliding electrons.

 {\it Before the collision} states 1a and 2a are occupied. The circle
 representing state 1a is below the dashed line, i.e. below the
 Fermi level $\mu$. The circle 2a represents a state with $p>0$,
 which is also occupied as the corresponding energy is below
 $\mu^{(+)}$.

 {\it After the collision} state 1b is occupied. It is
 represented by a circle above the dashed line, which means that
 it has been free before the collision. In wire 2 state 2b with
 $p<0$ is also occupied. It is above $\mu^{(+)}$, i.e. it had
 been free before the transition.

 The width of the stripe between the two straight lines is $eV$.
 If the bottoms of the active subbands are well below the Fermi
 level the drag current should be proportional to the number of
 the occupied initial states as well as to the number of free
 final states.

 To calculate the drag current, one can recast the product of
 distribution functions in the collision term Eqs. (\ref{B2}) and
 (\ref{3a}) into the form
 \begin{eqnarray}\label{ct_2}
 {\cal P}=2\sinh{\left({eV/{2T}}\right)}{\cal Q}
 \end{eqnarray}
 where
 \bean
 \label{4f}
 {\cal Q}&=&\exp{\varepsilon^{(1)}_{nk}-\mu\over T}
 \exp{\varepsilon^{(2)}_{n\prime k\prime}-\mu\over T}
 f(\varepsilon^{(1)}_{nk}-\mu)f\left(\varepsilon^{(2)}_
 {n^{\prime}k^{\prime}}-\mu^{(+)}\right) \nn\\
 & & \quad \quad \times f(\varepsilon^{(1)}_{nk^{\prime}}-\mu)
 f\left(\varepsilon^{(2)}_{n^{\prime}k}-\mu^{(-)}\right).
 \eean
 For the drag current one gets
 \ben
 \label{drag_cur}
 I= -\sinh{\left({eV\over2T}\right)}
 {8e^5mL\over{\pi^2\kappa^2}} 
  \times \sum_{nn^{\prime}}\int_0^{\infty}dk\int_0^{\infty}dk^{\prime}
 \displaystyle{g_{nn^{\prime}}(k+k^{\prime})\over{k+k^{\prime}}}
 {\cal Q}.
 \een

 As above, one can conclude that the terms which give the main
 contribution to the drag current are those where
 $\vert\tilde{\varepsilon}^{(1)}_n(0)-\tilde{\varepsilon}^{(2)}_{n'}(0)
 \vert$ is smaller than or of the order of $k_{\rm B}T$ or $eV$. We will
 assume that there is only one such difference (otherwise we would have
 got a sum of several terms of the same structure).

 As $\cal Q$ is a sharp function of $k$ and $k'$, one
 can take out of the integral all the slowly varying
 functions and get (the result is given for the general case where
 $\mu^{(1)}\neq\mu^{(2)}$)
 \ben
 \label{part_case}
 I= I_0\cdot{1\over2}\sinh\left({eV\over2k_{\rm B}T}\right)
 \frac{\displaystyle{eV\over4k_{\rm B}T}-{\tilde{\varepsilon}_{nn'}
 \over 2k_{\rm B}T}}
 {\sinh\left(\displaystyle{eV\over4k_{\rm B}T}-
 {\tilde{\varepsilon}_{nn'}\over2k_{\rm B}T}\right)}
 \times \frac{\displaystyle{eV\over4k_{\rm B}T}+{\tilde{\varepsilon}_{nn'}
 \over 2k_{\rm B}T}} 
  {\sinh\left(\displaystyle{eV\over4k_{\rm 
 B}T}+{\tilde{\varepsilon}_{nn'}\over
 2k_{\rm B}T}\right)}
 \een
 where
 \begin{equation}
 I_0=-{64e^5m^3L(k_{\rm B}T)^2\over{\kappa^2\pi^2\hbar^4}}
 \displaystyle{\cal D}_{nn'}.
 \end{equation}
 Here
 $\tilde{\varepsilon}_{nn^{\prime}}=
 \tilde{\varepsilon}^{(1)}_{n}(0)-
 \tilde{\varepsilon}^{(2)}_{n^{\prime}}(0).
 $

 For $eV\ll k_{\rm B}T$ Eq. (\ref{part_case}) turns into Eq.
 (\ref{19''}). Let us consider the opposite case $eV\gg k_{\rm B}T$.
 One gets for the drag current
 \begin{equation}
 I={\cal B}
 \left[\left({eV\over2}\right)^2-
 \left({\tilde{\varepsilon}_{nn'}}\right)^2\right],\quad
 {\cal B}=-{16e^5m^3L\over{\kappa^2\pi^2\hbar^4}}
 \cdot\displaystyle{\cal D}_{nn'}.
 \label{2}
 \end{equation}
 This result is nonvanishing only if
 $|\tilde{\varepsilon}_{nn\prime}|\,<\,eV/2$.

 In this Section we have discussed a Fermi liquid theory of
 the Coulomb drag current in a quantum wire brought about by a
 current in a nearby parallel quantum wire. A ballistic transport in
 both quantum wires is assumed. The drag current $I_D$ as a function of
 the wire widths comprises one or several spikes; the position of
 each spike is determined by a coincidence of a pair of levels of
 transverse quantization, $\varepsilon_{n}(0)$ and
 $\varepsilon_{n'}(0)$ in both wires.

 \begin{figure}[htb] 
   \begin{center}
     \epsfxsize 8cm
   \end{center}
   \caption{Schematic representation (for $\mu^{(1)}=\mu^{(2)}$) of
     simultaneous transitions due to the interaction between electrons of
     the two wires for $eV\gg k_{\rm B}T$. Circles $\circ$ and $\bullet$
     represent the initially unoccupied and occupied states respectively
     }
   \label{fig-2}
 \end{figure}

 \begin{figure}[htb] 
   \begin{center}
     \epsfxsize 8cm
   \end{center}
   \caption{ $I/I_{D}$ is plotted (for $\mu^{(1)}=\mu^{(2)}=\mu$) as a
     function of $W_1/W_2$ where the width of wire 1 is controlled through
     gate voltage ($\mu=$\,14\,meV, $T=$\,1 Kelvin, $W_2=$\,42\,nm,
     $L$=\,1\,$\mu$m, $\kappa\,$=\,13 and the spacing between wires is
     50\,nm).
     }
   \label{fig-3}
 \end{figure}

\section{Luttinger Liquid Theory of Coulomb Drag}\label{sec-luttinger}

 In 1D systems e-e interaction gives rise
 to electronic correlations that are believed
 to destroy the Fermi liquid. Instead, a different
 state is generated that is usually described as a Luttinger liquid 
 \cite{tomonaga__luttinger,haldane} (for reviews see e.g.
 \cite{voit,emery,solyom,schulz}).
 It is therefore not surprising that in  1D systems 
 e-e interaction affects the drag in a different
 way than in two- or three dimensional systems. 
 Indeed, in 1D systems interaction strongly
 enhances the effect, getting stronger the lower the temperature.
 As a result, the positive temperature characteristic of the drag
 resistance from Fermi liquid theory can become a negative
 one. For sufficiently long wires the drag resistance becomes 
 exponentially large at low temperatures.

 This chapter reviews in the main part the works
 \cite{nazarov,ponomarenko,klesse_stern}, and is organized as follows:
 Sec.~\ref{bosonization} introduces bosonic variables as 
 the appropriate language for the discussion to follow.
 In Sec.~\ref{drag} the renormalization group method is employed 
 in order to show that and in which way the drag becomes enhanced by
 electron correlations. This consideration 
 will also clarify some relevant energy and
 length scales of the problem. Sec.~\ref{spin}
 elaborates on the influence of the electron spin on the drag, while
 Sec.~\ref{non-linear} deals with non-linearities and 
 asymmetric double-wires. The last section of this chapter,
 Sec.~\ref{miscellaneous}, briefly discusses the drag 
 in a system with a finite region of interaction.

\subsection{Bosonic variables}\label{bosonization}
 For treating interactions it is convenient to describe
 excitations of the many-electron system in terms of collective
 coordinates: for example by the displacement $\fie(x)$ of electrons.
 They are normalized in such a way that density and current fluctuations
 are given by $ \partial_x \fie(x) = -\sqrt{\pi} (n(x) -n_0)$ and
 $ \partial_t \fie(x) = \sqrt{\pi} I(x)/e$. 
 Rewriting the Hamilton of an interacting 
 1D electron gas in $\fie(x)$ and its canonical conjugated field
 $\Pi(x)$ yields the Hamiltonian of an elastic string
 \ben\label{single_wire}
 H = \fr{v}{2} \int dx K \Pi^2 + \fr{1}{K} (\partial_x \fie)^2\:.
 \een 
 The stiffness or interaction parameter $K$ and the velocity $v$ 
 are determined by the parameters of the electronic system. 
 For non-interacting electrons $K=1$, $v=v_F$, while for
 a system with repulsive interaction $0 <K < 1$ and $ v \approx v_F/ K $.
 The solutions $\fie(x,t)$ of the Hamiltonian \Ref{single_wire}
 are 1D waves with wave velocity $v$. In the limit of
 strong interactions $K\ll 1$, these solutions correspond to the plasma
 oscillations of the electron density. Contrary to the 
 underlying fermionic operators, the fields $\fie$ and $\Pi$ obey 
 bosonic commutation relations. The substitution of the
 former by the latter is therefore known under the name of
 "bosonization" \cite{haldane,voit,emery,solyom,schulz}.

 Excitations of a double wire can be similarly described
 by the respective displacement fields $\fie_1(x)$ and $\fie_2(x)$ of
 each wire. Assuming a symmetrical system, the two eigenmodes of 
 the density oscillations (at a given wavenumber $q$) are the symmetric
 mode ($+$), where the density in both wires oscillate in phase, and the
 anti-symmetric mode ($-$), where the phases of the density oscillations differ
 by $\pi$. A Transformation to the corresponding displacement fields
  $\phi_\pm = (\fie_1 \pm \fie_2)/\sqrt{2}$ decouples the Hamiltonian into  
 a symmetric and anti-symmetric part,
 \bea
  H &=& H_1 + H_2, \\
 H_\pm &=& \fr{v_\pm}{2} \int dx \: K_\pm \Pi^2 + \fr{1}{K_\pm} (\partial_x
 \phi_\pm)^2\:.
 \eea
 Each part has its own set of parameters. In good approximation 
 \ben\label{Kspinless}
 K_\pm = \left( 1 + \fr{V_0 \pm \bar V_0}{ \pi v_F} -
   \fr{V_{2k_F}}{\pi v_F} \right)^{-1/2}\:,
 \een
 and $ v_\pm = v_F / K_\pm $,
 where $ V_0, \bar V_0$ are the Fourier transforms of
 intra-wire and inter-wire interaction $V(x)$ and $\bar V(x)$ for 
 small momentum $q\to 0$. $V_{2k_F}$ is the intrawire backscattering 
 strength ($\delta q = 2k_F$) \cite{klesse_stern}. 
 For the range of applicability of expression \Ref{Kspinless} see
 Refs. \cite{creffield,haeusler}, where more precise estimates of 
 the Luttinger-liquid parameters are given.

 So far interwire backscattering of electrons, where a large momentum of
 order $\delta q = 2 k_F$ is exchanged, has not been taken into
 account. As pointed out in the previous chapter, 
 this coupling however is essential for the drag and must be
 incorporated in the description. 
 Fortunately, this can be also done
 in terms of the displacement field, leading to
 \ben\label{H_back}
 H_b =  \lambda \fr{E_0^2}{\pi v_F} \int dx \cos( \sqrt{8\pi}
 \phi_-)\:.
 \een
 The energy $E_0$ is of order of the Fermi energy, and the
 dimensionless coupling $\lambda$ is given by
 \ben\label{lambda_0}
 \lambda = \fr{ \bar V_{2 k_F} }{ 2\pi v_-}\:,
 \een
 Note that the symmetric and anti--symmetric modes remain still decoupled.
 There is no corresponding term in the intra-wire interaction.
 The reason is that the backscattering within a wire appears as 
 the exchange part of the forward scattering ($\delta q\to 0$), and
 therefore can be absorbed in the parameters $K_\pm$ and $v_\pm$
 (cf. Eq. \Ref{Kspinless}).

 The origin of the backscattering Hamiltonian $H_b$
 becomes clear in the limit of strong repulsive
 intra--wire interaction. In this case the electrons of
 each wire form well-correlated states with charge densities periodic
 in $2\pi / k_F$. Accordingly, their local interaction energy is
 $2\pi$--periodic in the relative displacement 
 $(s_1 - s_2) k_F =  \sqrt{8 \pi} \phi_-$ 
 The integral over $\cos{\sqrt{8 \pi}   \phi_-(x)}$ with an appropriate
 prefactor therefore gives to first order the corresponding part of the
 total energy. 

\subsection{Drag}\label{drag}
 The backscattering Hamiltonian \Ref{H_back} is of sine-Gordon type, and 
 allows for an intuitive understanding of the drag in the case of large
 couplings $\lambda$. Suppose that the total energy is dominated by
 $H_b$, the system minimizes its energy by fixing the field $\phi_-$ to
 a value $\sqrt{8 \pi} \phi_0 = \pi + 2\pi m$, $m$ an integer
 number.  Accordingly, the relative displacement of electrons in
 wire 1 and 2 is constant in time and space. 
 This means that two interlocked
 charge density waves have been formed, such that a current in one wire is
 necessarily accompanied by an equally large current in the second
 wire. In this ideal situation the drag is absolute \cite{nazarov}.

 What happens to the drag if the situation is not that ideal is the 
 issue of this section. Considerable insight with a minimum of 
 calculation effort will be gained by making use of the concept of
 renormalization.  

 Neglecting for a while all interactions except for the inter-wire
 backscattering,  the double wire system can be viewed as a pair of
 uncorrelated 1D Fermi liquids. As explained in the previous section,  
 the inter-wire backscattering coupling then causes a drag
 resistivity $\rho_D = R_D/L$, where $L$ is the length of the drag
 wire,  linear in temperature and proportional to $\lambda^2$, 
 \ben\label{simple_drag}
 \rho_D \approx \rho_0 \lambda^2 T / E_0
 \een
 (cf. Eq. \ref{19''} in the limit of $T \gg \Delta \eps_n(0) $).
 To first order the drag only depends on the direct
 backward scattering part of the interaction. 
 However, higher order
 contributions to $\rho_D$ include inter- and also intra-wire forward
 scattering. In 1D these higher order contributions are
 crucial and must be taken into account. The renormalization group
 theory does the job quite elegantly by successively integrating out
 high energy degrees of freedom down to an energy scale $E < E_0$. As a
 result, the original ("bare") couplings $K_-$, $\lambda$
 become renormalized to $E$--dependent couplings $K_-(E)$ and
 $\lambda(E)$. The energy scale at which the renormalization procedure
 has to be stopped can be given either by the temperature, the
 system size, or even by the coupling $\lambda(E)$ itself, depending
 on the circumstances.
 The net effect of higher order processes on the
 drag between a pair of weakly coupled wire  can be summarized by
 replacing -- for example in Eq. \Ref{simple_drag} -- the bare coupling $\lambda$
 by a
 renormalized energy dependent ("running") coupling constant
 $\lambda(E)$.  
 Further, good approximations for relevant energy scales can be easily
 extracted from the renormalization procedure. This should be enough
 motivation for a short excursion to the renormalization flow of 
 the sine-Gordon model. 

\subsubsection{Renormalization flow}
 The flow is well-known from a closely related problem, that of an interacting
 spin-$1/2$ electron liquid \cite{spin1/2}. For small couplings $\lambda \ll 1$
 it 
 is described by the differential equations
 \ben\label{flow}
 \fr{d \lambda}{dt} = (2- 2 K) \lambda, \quad 
 \fr{d K}{dt} = -2 \lambda^2 K^2,
 \een
 where $t$ denotes the negative logarithm of the rescaled energy,
 $t  = \ln E_0 / E  $. (The subscript ``$-$'' is
 suppressed henceforth.) Fig. \ref{fig-flow} shows schematically the
 flow in a $K-\lambda$--diagram. Each point represents a system 
 characterized by the parameters $K$ and $\lambda$. Under
 renormalization the system develops according to the 
 stream lines in the parameter space. The arrows indicate the direction
 of decreasing energy scale. 
 \begin{figure}[htb] 
   \begin{center}
   \end{center}
   \caption{ The RG-flow of a double wire system of spin-less
     electrons. Point A corresponds  to the bare couplings of a
     double wire with a rather large inter-wire distance  $ d  \gg \lambda_F $, 
     point B to wires with narrow spacing $ d \ll \lambda_F $. Point C corresponds
     to a single spin-1/2-liquid.
     }
   \label{fig-flow}
 \end{figure}
 
 The main feature of the flow is the so--called Kosterlitz-Thouless
 transition: systems with parameter below the line $K=1+\lambda$ renormalize
 towards weaker backward scattering $\lambda$, systems above or left of
 this line renormalize towards larger $\lambda$. Systems
 right on the transition line flow to the point at $K=1$ and
 $\lambda=0$, which represents a non-interacting Fermi liquid.

 Whether a system develops to weaker or stronger couplings
 $\lambda$ obviously depends on the location of the bare 
 coupling (subscript ``0''),
 the initial points of the renormalization trajectories. 
 For systems with symmetrical interaction, $V(x)  = \bar V(x)$,
 as it is the case for a spin-$1/2$ electron liquid, one
 finds for small $V_{2k_F} $ that $K_0 = 1 + \lambda_0$, i.e. the
 initial point lies exactly on the transition line. These systems
 (marginally) renormalize towards the non-interacting Fermi-point.
 This is indeed the expected behaviour for the spin-mode of an electron
 liquid. 
 (The spin-mode corresponds to the anti-symmetric mode ($-$).)
 If one could measure the relative drag of spin-up and spin-down
 electrons, one would find that due to the additional temperature
 dependence in $\lambda(T)$, the drag resistivity decays with 
 temperature even faster than the naively expected linear behaviour.

 Interestingly, the situation is completely different for a real
 double-wire system. Due to the spatial separation of the two wires,
 the intra-wire interaction  $V(x)$ always exceeds the inter-wire 
 interaction $\bar V(x)$. Inspection of Eq.s \Ref{Kspinless} and
 \Ref{lambda_0} reveals that as a consequence the initial point is 
 always above or left of the transition line, where $\lambda$
 renormalizes to higher values. Therefore here the
 backscattering coupling $\lambda$, although usually much smaller than
 in the previous case, increases with decreasing energy scale or
 temperature. Hence, for low temperatures (and long wires, see below)
 the drag resistivity will be always larger than predicted by Fermi
 theory. This will be made more quantitatively in the next paragraph.

\subsubsection{Temperature dependence of the drag}
 Before rushing into a discussion of the various kinds of regimes with
 different types of temperature dependencies, it is advisable to clarify 
 the relevant energy and length scales. There is 
 $E_0 \sim E_F$, the largest energy scale, the temperature 
 $T_L$, at which the thermal wavelength becomes of the order of the
 system size, and the actual temperature $T$. 
 The corresponding length scales are the Fermi wavelength $\lambda_F
 \sim E_0/ v $, the system size $L$, and the thermal wavelength $L_T =
 v/T$. Less obviously, a fourth energy and length scale are given
 by the sine-Gordon Hamiltonian $H_- + H_b$: the energy (mass) $M$ of a
 soliton and its width $L_s$. The soliton mass $M$ coincides with the 
 energy scale at which the renormalization procedure breaks down, the 
 soliton width $L_s$ is the corresponding length scale. 
 The relative order of these scales classify several different regimes.

 For high temperatures $T > T_L, M$, the renormalization of
 $\lambda$ is terminated by $T$. Thus, $\lambda = \lambda(T)$.
 In this case the temperature dependence of $\lambda(T)$ can be approximately 
 determined by integrating the flow equations \Ref{flow}, leading to
 \ben\label{lambda_T}
 \lambda(T) = \lambda_0 \left(\fr{T}{E_0}\right)^{2K-2}\:.
 \een
 This result inserted in Eq. \Ref{simple_drag} gives the temperature
 dependence,
 \ben\label{real_drag}
 \rho_D \approx \rho_0 \lambda_0^2  \left( \fr{T}{E_0} \right)^{4K -3}\:,
 \een
 valid for $T_L, M \ll T \ll E_0$.
 The e-e interaction changes via the renormalization
 of the backscattering coupling, the linear temperature dependence
 known from the 1D Fermi-liquid to a temperature scaling with
 an interaction dependent power $\chi= 4K-3$.
 For sufficiently strong interaction $K$ can assume values below $3/4$.
 Then the power $\chi$ becomes negative and the drag increases
 with decreasing  temperature. For vanishing interaction, $K=1$,
 Eq. \Ref{real_drag} goes over to the linear behaviour of the
 Fermi-liquid.

 Lowering the temperature below $M$ or $T_L$, two scenarios are
 possible: if the wire is sufficiently long, $L \gg W$, at 
 a temperature $T\sim M$ the system eventually enters the strongly
 coupled regime, where $\lambda(T) \sim 1$.
 For short wires, $L \lapprox L_s$, this regime can not be reached.
 Here the renormalization halts at a temperature $T\sim T_L$,
 where the thermal wavelength is of order the system size. In this
 case even at low temperatures the systems are weakly coupled,
 $\lambda(T_L) \ll 1$.

 In the strongly coupled regime the energy is dominated by the
 backscattering term $H_b$, giving rise to an almost absolute drag.
 Deviations from this ideal drag correspond to processes where the
 relative displacement $\sqrt{8 \pi} \phi_-$ slips from one global
 minimum position, say at $\pi$, to a neighbouring one at $-\pi$ or $3 \pi$.
 At temperatures $T < M$ these processes are enabled by thermally
 activated solitons moving along the wire. As a result the drag
 resistivity shows for $T<M$ an activated behaviour, 
 \be
 \rho_D(T) \sim \tilde\rho_0 e^{M/T}\:.
 \ee
 This behaviour changes again when the temperature falls below $T_L$
 (for the strong coupling regime considered, $T_L < M$). It has been shown
 \cite{ponomarenko} that then the drag decreases linearly with
 temperature, due to the set-in of coherent soliton-tunneling. At even lower 
 temperatures $T < \sqrt{T_L M} \exp{-M/T_L}$ the drag decreases with
 $T^2$ \cite{ponomarenko}.

 In the weakly coupled regime, the drag resistance decays $\propto T^2$
 as the temperature drops below $T_L$. This can be understood as
 follows: having renormalized down to an energy $T_L$, the original 
 Fermi wavelength length $\lambda_F \sim v/E_0$ has become enlarged to
 a rescaled wavelength $\lambda_F(T_L) \sim v/T_L = L$. Hence, the
 wires of length $L$ have effectively shrunk down to pointlike
 constrictions connecting electronic reservoirs on either side. The
 drag in this situation is equivalent to the drag in a pair of 1D
 Fermi-liquids ($K=1$) that interact over a short length $\lapprox
 L_T$ only. As it will become clear in the Sec. \ref{miscellaneous},
 the drag is then proportional to $T^2$ (cf. Sec. \ref{miscellaneous}, $K=1$).

 The temperature scale $T_*$ at which the system enters the strongly coupled
 regime is
 given by the soliton mass $T_* = M$. Estimating it by  
 $\lambda(T_*) \sim 1$ with the approximate expression
 \Ref{lambda_T} yields
 \ben\label{Tstarspinless}
  T_* \sim E_0 \: \lambda^\fr{1}{2 - 2K}\:.
 \een
 The minimum wire length required is then $L_* = v/T_* = \lambda_F
 \lambda^{-\fr{1}{2-2K}}$. 

\subsection{Electron Spin}\label{spin}
 For comparison to experiments the treatment of the electron spin 
 is mandatory. To this end one can introduce bosonic
 fields $\fie_{c/s}$ that are related to the charge/spin density $n_c =
 n_\uparrow \pm  n_\downarrow $ in the same way as above $\fie$ is related to the
 density $n$ of spinless particles. For a double wire, this results
 in a total of four modes: symmetric and anti-symmetric charge
 modes ($c+$ and $c-$) as before, and, additionally, symmetric and anti-symmetric
 spin modes ($s+$ and $s-$). Each mode is again described by a
 quadratic Hamiltonian of the type \Ref{single_wire} with corresponding 
 interaction parameters $K_{c\pm}, K_{s\pm}$, etc. . 
 The neutral spin-modes are not affected by the interaction, wherefore
 $K_{s\pm} = 1$ and $v_{s\pm} = v_F$. Nevertheless, despite their neutrality, the 
 spin-modes weakly couple to the anti-symmetric charge
 mode $c-$ via backscattering processes. Hence, the drag is influenced by  
 the spin degree of freedom \cite{klesse_stern}. 

 A quantitative analysis of the weakly coupled regime
 can again be done by making use of  renormalization along the lines
 described above. The main results are summarized below: 
 In the presence of spin the interwire backscattering coupling 
 scales towards stronger couplings. However, fluctuations in the neutral
 spin-modes moderate the enhancement due to the interactions. This is reflected 
 by an effective interaction parameter
 \be
 K_{eff} = \fr{ K_{c-} + K_{s\pm} }{2} = \fr{K_{c-}+1}{2}\:
 \ee
 which is closer to the non-interacting value $1$ than the original $K_{c-}$.
 The parameter $K_{c-}$ is given by
 \ben\label{Kspinfull}
 K_{c-} \approx \left( 1 + 2\fr{V_0 - \bar V_0}{ \pi v_F} -
   \fr{V_{2k_F}}{\pi v_F} \right)^{-1/2} \:.
 \een
 As a result, in the weakly coupled regime the drag resistance 
 scales with temperature as
 \be
 \rho_D \approx \rho_0 \lambda_0^2 
 \left( \fr{ T}{E_0} \right)^{2K_{c-} -1}\:
 \ee
 (cf. Eq. \Ref{real_drag}). The cross-over temperature
 $T_*$ turns out to be approximately
 \be
 T_* \sim E_0 \: \lambda^{\fr{1}{1-K_{c-}}}\:.
 \ee
 Comparison with Eq. \Ref{Tstarspinless} reveals again the moderating
 effect of the spin. If two systems have similar interaction constants
 $K_- \approx K_{c-}$, but one is spin-polarized while the other is
 not, their respective cross-over temperatures and lengths are related
 by 
 \be
 \left( \fr{T_*}{E_0} \right)^2_{pol.}  \approx 
 \left( \fr{T_*}{E_0} \right)_{un-pol.}, \quad 
 \left( \fr{\lambda_F}{L_*} \right)^2_{pol.}  \approx 
 \left( \fr{\lambda_F}{L_*} \right)_{un-pol.}\:.
 \ee
 Since $T_*/ E_0 \sim \lambda_F/L_*$ is usually a small number, the
 cross-over temperature of the spin-unpolarized system is by orders of
 magnitudes smaller than the one of a comparable spin-polarized system.

\subsection{Non-linear drag and mismatching Fermi-momenta}\label{non-linear}
 So far our considerations were confined to the linear regime $(I\to
 0)$ of a symmetrical double wire system. This section extends the
 discussion to both the non-linear regime, and to systems with a misfit
 in the Fermi momenta, $\delta k = k_{F1} - k_{F2} \neq 0$.  

 It is again useful to look first at the associated energies.
 A finite current $I$ in the active wire defines an energy
 $\Omega = I/e$, and the energy associated to the misfit is of course
 $\Delta = v \delta k$. 
 Non-linearities of the drag voltage $V_D$ in the current $I$, or an
 effect of the misfit $\delta k$ will be significant only if the
 corresponding energies $|\Omega|$ or $|\Delta|$ exceed $T$, $T_L$ and
 $M$. 


 In the weakly coupled regime both cases can be analyzed by perturbative 
 methods. Finite currents and a non-vanishing $\delta k$ can
 be treated by a transformation $\phi(x,t) \to \phi(x,t) + \Delta x /
 v + \Omega t$. The term linear in $x$ describes the density difference
 ($\partial_x \phi \propto (n_1- n_2)$), the term linear in $t$
 corresponds to a Galilei boost of the active wire. Accordingly, $H_b$
 becomes 
 \be
 H_b = \lambda \fr{E_0^2}{\pi v_F} \int dx \: \cos \sqrt{8\pi}( \phi +
 \Delta x/v + \Omega t)\:.
 \ee
 It is possible to derive a closed expression for the drag voltage
 $V_D$ as a response to this perturbation \cite{unpublished}. It is valid 
 for arbitrary ratios $\Omega/\Delta$, and can be written as
 \bean\label{dragvoltage}
 \fr{eV_D}{ L } &=& C \fr{E_0^2 \lambda_0^2 }{v} \left( \fr{T}{E_0}
 \right)^{4K}  \{ A(\Omega - \Delta) B(\Omega + \Delta)  \nn \\
 & & + A(\Omega + \Delta) B(\Omega - \Delta) \}\:.
 \eean
 $C$ is a numerical constant of order unity, $A$ and  $B$ are 
 temperature dependent functions, given by
 \be
 A(E) = \int ds \: i \sin(\fr{E}{E_0} s) \left( \pi (\fr{1}{s} + i)
   \sinh\fr{Ts}{\pi E_0 } \right)^{-2K}\:,
 \ee
 and a similar expression with $\cos$ instead of $i \sin$ for $B$.
 This expression holds also for vanishing $\Delta, \Omega$, where it
 leads to  the  result \Ref{real_drag}. For current and $\delta k$
 large compared to temperature, $\Omega, \Delta \gg T$,
 Eq. \Ref{dragvoltage} reduces to \cite{nazarov}
 \bea
 \fr{eV_D}{L} &=& const.\: \lambda^2_0 \:(\Omega^2 - \Delta^2)^{2K-1},
 \quad \mbox{ for } |\Delta| < |\Omega| \\
 &=& 0 \quad\qquad\quad\quad\quad \quad  \mbox{ otherwise } \:.
 \eea
 For non-vanishing $\Delta$ the drag vanishes as long as the current 
 is below a threshold value. For larger currents the voltage shows
 a powerlaw dependence on the current. A thorough discussion of the 
 non-linear  $I-V$ characteristic can be found in \cite{nazarov}.

 Actually, at finite temperatures the drag does not vanish completely
 for $|\Delta| > |\Omega|$, rather it shows an activated behaviour as 
 like in the case of 1D Fermi liquids. This can be made more explicit
 \cite{unpublished}.
 Evaluating expression \Ref{dragvoltage} in the limit $\Omega \to 0$ at
 finite $\Delta$ results in a drag resistivity
 \ben\label{activedrag}
 \rho_{\Delta,T} = \rho_0 \lambda^2 \left( \fr{T}{E_0} \right)^{4K-3} 
 F_{2K}(\Delta/T)\:,
 \een
 where $F_{2K}$ is an interaction dependent function defined by
 \bea
 F_{2K}(\eps) &= & N_{2K} (\eps) \fr{dN_{2K}}{d \eps} ( -\eps ) + 
 N_{2K}(-\eps) \fr{d N_{2K}}{d \eps} ( \eps )\:, \nn \\
 N_{2K}(\eps) &=& \lim_{\delta \to 0} \int ds \: e^{i s\eps/\pi} 
 \left( ( \fr{\delta}{s} + i) \sinh s \right)^{-2K}\:.
 \eea
 $F_{2K}$ is a continuous function. It decays exponentially at
 large positive arguments, $F_{2K}(\eps) \sim \exp{-\eps}$, and behaves
 algebraically for large negative arguments, $F_{2K}(\eps) \sim |\eps|^{K-1}$.
 The result obtained for non-interacting 1D Fermi-liquids is
 recovered by putting $K=1$. In this case
 \be
 N_2(\eps) = \fr{1}{\pi} \fr{\eps}{e^\eps -1}\:,
 \ee
 such that the expression \Ref{activedrag} corresponds to Eq. 
 \Ref{19''}. While $N_2(\eps)$ bears some resemblance
 to the Bose distribution, for the special interaction parameter
 $K=1/2$ one obtains exactly the Fermi function
 \be
 N_1(\eps) = \fr{1}{e^\eps +1 }\:.
 \ee
 For general parameter $K$ an analytical expression for $N_{2K}$ is
 lacking. It is an open  question whether the functions $N_{2K}$ are
 related to  the exclusion statistics of fractional
 excitations in the Luttinger liquids. 

\subsection{Finite interaction region}\label{miscellaneous}
 Double wires that interact only over a region of finite length $l < L$
 have been also investigated \cite{komnik,flensberg}. For temperatures $T < v/l$
 this 
 problem can be mapped to the classical problem of a
 Luttinger liquid with a single impurity. Qualitatively, these systems
 behave similar to those considered in the previous sections. In the
 weakly coupled regime the drag scales with temperature 
 with an interaction dependent exponent, $4K - 2$. A strongly coupled
 regime with almost absolute drag at zero temperature exists as well. 
 However, it is reached only for sufficiently strong interaction $K <
 1/2$. For $K>1/2$ the interwire backscattering coupling renormalizes
 to weaker couplings, such that the drag vanishes for $T\to 0$.

\section{Experimental search for 1D Coulomb drag} \label{sec-exp}

 Although a fair amount of theoretical work has been available on Coulomb 
 drag between 1D electron systems, there has been a conspicuous absence of 
 experimental work. This may be attributed to two difficulties encountered in 
 measuring the 1D drag. First, since it is a very small effect, the drag 
 voltage usually has a very small magnitude and must be clearly distinguished 
 from spurious signals. Second, and perhaps the major difficulty, has been 
 the difficulty in creating parallel, electrically isolated, quantum wires 
 with a spatial separation large enough to completely suppress interwire, 
 while small enough to give a drag voltage of a reasonable magnitude. It was 
 only recently that Debray \textit{et al} 
 \cite{debray2000} reported the first experimental observation 
 of Coulomb drag between ballistic quantum wires. The same authors later 
 published a more comprehensive experimental work \cite{debray2001} on the
 subject. Work 
 along the same lines has lately been reported by Yamamoto et al \cite{yamamoto}.
 In the 
 following, we give a brief outline of the reported experimental work and an 
 analytical discussion of the results in the framework of the Fermi and the 
 Luttinger liquid theory as discussed in Secs. II and III. 

\subsection{A. Experimental techniques for dual-wire sample
   realization} 
 The samples used for 1D Coulomb drag measurements consisted of two 
 electrically isolated, parallel quantum wires, with a small spatial 
 separation. Such samples were fabricated from AlGaAs/GaAs heterostructures 
 with a high-mobility ($ \cong $10$^{{\rm 6}}$cm$^{{\rm 2}}$/Vs) 
 two-dimensional electron gas (2DEG) at the interface. The dual-wire samples 
 were fabricated using high-resolution electron beam lithography, combined 
 with deep chemical etching. The samples used so far were made in a planar 
 geometry by depletion of a single 2DEG layer by three surface Schottky gates 
 \cite{debray2000,debray2001,yamamoto} deposited on the
 heterostructure wafer. Figure \ref{fig-5} gives a schematic top view
 of the planar device and the scanning electron micrograph of a
 typical device \cite{debray2001}. U, M, and L are surface Schottky
 gates. 
 \begin{figure}[htb] 
  \begin{center}
          \epsfxsize 6cm
  \end{center}
  \caption{ (a) Schematic top view of a planar Coulomb drag device. U, 
    M, and L are surface Schottky gates. (b) A scanning electron micrograph of a 
    typical device with middle gate width of 50 nm.}
  \label{fig-5}
 \end{figure}

 Dual-wire samples for drag measurements can also be made in a vertical 
 geometry \cite{moon} from two vertically stacked quantum well (QW) structures
 with 
 a 2DEG in each well (such samples have not yet been used). The advantage of 
 the planar geometry is that the interwire separation can be changed in-situ 
 by changing the bias voltage of the central gate M. The disadvantage is that 
 the narrow central gate creates a soft lateral potential barrier and to 
 prevent tunneling between the wires the width of this barrier has to be of a 
 sufficient magnitude, which sets a limit to the minimum interwire distance 
 that can be used without electron tunneling interfering. Samples with a 
 vertical geometry have been widely used for studying Coulomb drag between 2D 
 electron layers [2]. The main advantage of the vertical geometry is that 
 very small interwire separation (the barrier width) can be obtained without 
 tunneling between wires. Since the magnitude of the drag is expected to 
 decrease exponentially with interwire separation, one can expect to observe 
 enhanced drag with the vertical samples because of the smaller separation 
 that can be achieved with such samples. The major disadvantage is that the 
 interwire separation cannot be changed in-situ, in contrast to the planar 
 case. Also, it is not obvious that the widths of the two wires can be 
 independently changed through the use of the mutually aligned top and bottom 
 split gates.

\subsection{B. Experimental observation of Coulomb drag}

 In their work, Debray et al \cite{debray2001} used a planar geometry
 of quantum wires, as  
 shown in Fig. 5, of lithographic length $L $= 2 $\mu $m with a middle gate M of 
 lithographic width 50 nm. The drag voltage $V_{D}$ was measured with a drive 
 voltage V$_{{\rm D}{\rm S}}$ in the linear\textit{} regime\textit{} of
 ballistic\textit{} electron 
 transport as a function of the width of the drive wire by adjusting the bias 
 voltage V$_{{\rm U}}$, while the width of the drag wire was adjusted to have 
 the Fermi level $E_{F}$ just above the bottom of its lowest 1D subband. An 
 appropriate negative bias voltage V$_{{\rm M}}$\textit{} was applied to the
 middle 
 gate to ensure total absence of interwire tunneling. Measurements were done 
 in the\textit{ absence} of any such tunneling. In Fig. \ref{fig-6}a is shown
 the measured drag voltage   
 $V_{D}$ as a function of the width of the drive wire. The drag voltage is 
 found to show peaks, which occur in the rising parts between the plateaus of 
 the drive wire conductance. This suggests that they occur when the 1D 
 subbands of the wires are aligned and the Fermi wave vector k$_{{\rm F}}$ is 
 small. Measurements carried out in a magnetic field B = 0.86 Tesla 
 perpendicular to the plane of the device shown in Fig.6(b) indicate 
 identical behavior except that the magnitude of $V_{D}$ is enhanced almost by 
 a factor of three. 

  \begin{figure}[htb] 
  \begin{center}
          \epsfxsize 12cm
  \end{center}
  \caption{ The drag voltage $V_{D}$ and drive current $I$ as function
    of drive wire width at a drive voltage V$_{{\rm D}{\rm S}}$ = 300
    $\mu $V \cite{debray2001}. (a) In zero magnetic field with the
    upper wire as the drive wire. (b) In a magnetic field of 0.86 Tesla
    with the bottom wire as the drive wire.  }
  \label{fig-6}
  \end{figure}

 In order to have unambiguous evidence that the observed drag voltage 
 $V_{D}$ is indeed due to the Coulomb drag effect, the authors measured the 
 dependence of $V_{D}$\textit{} and $R_{D}$\textit{} on the interwire separation
 and the 
 temperature. Figure \ref{fig-7} shows the dependence of $V_{DM}$, the height of
 the 
 first $V_{D}$ peak of Fig. \ref{fig-6}b and the corresponding $R_{D}$ as
 function of 
 the middle gate bias voltage V$_{{\rm M}}$. The two quantum wires (Fig.
 \ref{fig-5}) 
 were spatially separated by an effective distance $d$ due to the depletion by 
 V$_{{\rm M}}$ of the 2DEG under the middle gate M. In the voltage range of 
 interest, $d$ was experimentally found to vary almost linearly with V$_{{\rm 
 M}}$ according to
 \begin{equation}
 \label{eq1}
 d = d_{0} + \alpha \left( {V_{0} - V_{M}}  \right){\rm ,}
 \end{equation}
 \noindent
 where V$_{{\rm 0}}$ is the value of V$_{{\rm M}}$ for which the 2DEG under M 
 is just depleted and $\alpha $ gives the total spatial displacement of the 
 two depletion edges of M with respect to its bias voltage. $d_{0}$\textit{} is a 
 constant for the same device and is nominally equal to the lithographic 
 width of the gate M. One can change $d$ by varying V$_{{\rm M}}$. In
 Eq.(\ref{eq1}), 
 V$_{{\rm 0}}$ and $\alpha $ were determined experimentally. The dependence 
 of $R_{D}$ on V$_{{\rm M}}$ was found to be exponential and can be described 
 well by the relation, $R_{D} \propto e^{\beta V_{M}} $, where $\beta \cong 
 14.2(9)V^{ - 1}$.

  \begin{figure}[htb] 
  \begin{center}
          \epsfxsize 12cm
  \end{center}
  \caption{Dependence of the drag response on interwire separation $d$ 
 via the middle gate voltage V$_{{\rm M}}$ \cite{debray2001}. (a) The maximum
 $V_{DM}$ of 
 the first drag peak of Fig. 6(b) as a function of V$_{{\rm M}}$. (b) The 
 natural logarithm of the corresponding drag resistance $R_{D}$ as a function 
 of V$_{{\rm M}}$. The dotted line is a linear fit to the data points.   }
  \label{fig-7}
  \end{figure}

 The temperature dependence of Coulomb drag is a crucial feature that can be 
 used to probe which one of the two theoretical models, the FL or the LL 
 theory, constitutes a more appropriate description of 1D Coulomb drag. 
 Measurements, such as shown in Fig. \ref{fig-6}, carried out in the temperature
 range 
 60 mK - 1.2K are shown in Fig. \ref{fig-8}. A decrease of $V_{D}$ with
 increasing 
 temperature was observed. The dependence on temperature of the drag 
 resistance $R_{D}$ corresponding to $V_{DM}$ is shown in Fig. \ref{fig-9} for
 both in 
 the absence and presence of an applied magnetic field B. The temperature 
 dependence can be described well by the power law, $R_{D} \propto T^{x}$, 
 with $x$ = -0.77(2) and --0.73(6) for B = 0 and B = 0.86 Tesla, respectively. 
 It is interesting to note that the data points at temperatures lower than 
 180 mK, for zero field, and 300 mK, for nonzero field, fall below the 
 power-law curve, indicating a suppression of the drag effect.

  \begin{figure}[htb] 
  \begin{center}
          \epsfxsize 12cm
  \end{center}
  \caption{The dependence of drag voltage on temperature \cite{debray2001}. (a)
 The 
    drag voltage $V_{D}$ as a function of the width of the upper (drive) wire in 
    zero magnetic field with V$_{{\rm D}{\rm S}}$ = 300 $\mu $V at 70, 180, 300, 
    450, and 900 mK, corresponding to curves in order of decreasing peak height. 
    (b) The same as in (a) but in a magnetic field of 0.86 tesla with V$_{{\rm 
        D}{\rm S}}$ = 50 $\mu $V at 60, 180, 300, 450, 900 mK, and 1.2K. }
  \label{fig-8}
 \end{figure}

 \begin{figure}[htb] 
   \begin{center}
    \epsfxsize 12cm
  \end{center}
  \caption{The temperature dependence of drag resistance $R_{D}$ 
    corresponding to V$_{{\rm D}{\rm M}{\rm} }$of the first drag peak of Fig. 8 
    in zero field (a) and in a field of 0.86 Tesla (b)
    \cite{debray2001}. Note that the data  
    points at the low end of the temperature range fall below the
    power-law curve. }
  \label{fig-9} 
 \end{figure}

 Lately, using a lateral sample geometry , very similar to that shown
 in Fig. \ref{fig-5}a, Yamamoto et al \cite{yamamoto} has reported the
 observation of Coulomb drag and  
 the influence of an applied magnetic field on it. Their results corroborate 
 those of Debray et al \cite{debray2000} \cite{debray2001}. In their work,
 Yamamoto et al also reported 
 the observation of a negative drag. However, since the negative drag was 
 observed only when the drive wire was completely pinched off, it is highly 
 questionable if the effect observed is due to Coulomb drag. 

\subsection{C. Discussion}

 The origin of the observed peaks in the drag voltage $V_{D}$ (Fig.6) can be 
 understood when one considers Eqs. (16)-(21) of Sec. II. Since $V_{D}$ is 
 directly proportional to the drag current $I_{D}$, $V_{D}$ will show maxima 
 whenever any two 1D subband bottoms of the two wires line up and the Fermi 
 wave vectors in the two wires are equal and small. As seen from Fig. 6, the 
 occurrences of the drag peaks correspond to these conditions. The first peak 
 in $V_{D}$ occurs when the Fermi level is just above the bottoms of the 
 lowest 1D subbands of both wires. Similarly, the second peak occurs when the 
 lowest subband of the drag wire lines up with the second subband of the 
 drive wire. Both the increase and the narrowing of the first drag peak in a 
 magnetic field of 0.86 Tesla (Fig 6(b)) can be attributed to an increase of 
 the density of states in1D the subbands due to the magnetic-field-induced 
 enhancement of the electron effective mass. The reduction in the magnitude 
 of the drag peak as we move away from the first peak toward higher values of 
 V$_{{\rm U}}$ can be attributed to an increase in the effective interwire 
 separation of the wires. This dependence is explained in detail later. Since 
 we are mainly concerned with 1D transport in the fundamental mode, we 
 restrict our discussion here to the region of the first drag peak. 

 To understand the dependence of drag on the interwire distance shown in Fig. 
 7, we note that the matrix element of the backscattering probability depends 
 on the interwire distance $d$ via the modified Bessel function $K_{0} \left( 
 {2k_{F} d} \right)$ (Sec. II, Eq.(11)), which is an exponential function of 
 its argument for $2k_{F} d > > 1$. The same dependence also results from the 
 LL theory [29]. Under this condition, an exponential decrease of $R_{D}$ with 
 $d$ is expected according to $R_{D} \propto \exp \left( { - 4k{}_{F}d} 
 \right)$. This is consistent with the results of Fig.7. Using the 
 experimentally determined values $\beta $ = 14.2 V$^{{\rm -} {\rm 1}}$ and 
 V$_{{\rm 0}}$ = -0.4V, $\alpha $ = 580 nm V$^{{\rm -} {\rm 1}}$, we find 
 $k_{F}$ = 6.1 x 10$^{{\rm 6}}$ m$^{{\rm -} {\rm 1}}$. Surprisingly, this 
 corresponds to a low density of about 8 electrons per 2$\mu $m wire segment 
 and a mean electron distance 
 $\bar r \approx 250 nm$ in the wire. When V$_{{\rm M}}$ is in the range 
 from -0.7 to --0.8 V, we have (Eq. (\ref{eq1}) $d  \cong $ 0.2 $\mu $m. This
 gives 
 2$k_{F}d \quad  \cong $ 3, so the approximation of 2$k_{F}d \gg 1$ is
 reasonable. This exponential decrease of $R_{D}$ with $d$ also
 explains why the height of the  
 drag voltage peaks in Fig.6 decreases so rapidly as V$_{{\rm U}}$ increases. 
 An increase in V$_{{\rm U}}$ increases the width of the drive wire and hence 
 $d$. The decrease of $R_{D}$ for V$_{{\rm M}} > -0.7 V$ occurs due to tunneling 
 of a considerable fraction of the current from the drive wire to the drag 
 one, reducing the measured $R_{D}$.

 The experimental observed features of the drag effect discussed above, 
 namely, the origin of the drag voltage peaks, the effect the magnetic field, 
 and the interwire separation dependence, can all be understood in the 
 framework of both the FL and the LL theory. It is the temperature dependence 
 of the drag that is the crucial feature - it can be used to determine which 
 one of the two theoretical models constitutes a more appropriate description 
 of 1D Coulomb drag observed under the given experimental conditions. The 
 observed temperature dependence of $R_{D}$, shown in Fig.9, is in sharp 
 contrast with the linear temperature dependence predicted by the FL theory 
 (Eq. (19)). The unusual temperature dependence can not be attributed to a 
 temperature induced modification of the wire conductance, since the latter 
 is found to be almost unchanged over the temperature range of the 
 measurements. A reduction of the interwire Coulomb coupling due to enhanced 
 screening by the reservoirs and gates is very unlikely at such small 
 temperatures. On the other hand, it's conceivable that a correlated LL 
 behavior is established in the wires. Indeed, it is hardly surprising that 
 the temperature dependence of $R_{D}$\textit{} does not fit into a FL scenario,
 because 
 for the experimental condition of the first drag peak the ratio
 $r_{s}$ of $\bar r$ and the Bohr radius $a_{B}$, $ r_s = \bar r / a_B
 \approx 26$ is large. 

 The smallness of the drag resistance ($R_{D} < 100 \Omega $) in zero 
 magnetic field indicates a weak interwire back scattering coupling. In this 
 case, according to the LL model $R_{D}$ should obey a power law as long as 
 the thermal length $L_{T}$ is well in between the wire length $L$ = 2 $\mu $m 
 and the mean electron distance 
 $\bar r \approx 250nm$ in the wire. For spin-unpolarized electrons, valid 
 for the data shown in Fig.9, the LL description predicts a power-law 
 temperature dependence of $R_{D}$ with exponent 
 $ x = 2 K_{c-}-1$  (Eq.(38)). The data shown in Fig.9 indeed shows a 
 power-law dependence of $R_{D}$ on temperature with 
 $K_{c1} = 0.12$. Let us see if the condition 
 $\bar r < L_T < L$ is fulfilled in 
 the experiment. Given a Fermi wavevector of 
 $k_F \approx 6 \mu m \inv$
  we find that 
 $L_T = h v_F / K_{c-} k_B T$ is equal to the 
 wire length $L$ = 2 $\mu $m at a temperature $T_{L}  \cong $ 250 mK, and that 
 $L_{T}$\textit{}  approaches 
 $\bar r \approx 250nm$ at a temperature of about 2K. Here 
 $v_F / K_{c-}$
  is the group velocity of the relative 
 electron-density fluctuations and 
 $h/k_B T$ is the quantum lifetime associated with the thermal 
 energy 
 $k_B T$. This means 
 that there is a narrow temperature range window in which a power-law 
 temperature dependence of $R_{D}$ might be expected, and it is observed 
 experimentally. At temperatures below $T_{L}$, when $L<L_{T}, $the electron
 coming 
 from the lead to the wire does not have time to accommodate itself to the LL 
 liquid. This should result in a drag weaker than the power-law
 dependence.\textit{} The 
 experimental data of Fig.9 are consistent with this analysis. At lower 
 temperatures we do indeed observe a tendency to a weakening of the drag with 
 respect to the power-law dependence. 

 The negative power-law temperature dependence is not the only experimental 
 feature that can not be understood in terms of the FL theory of Coulomb 
 drag. The experimental value of $R_{D}$ (Fig. 9a) at T = 60 mK is more than 
 an order of magnitude larger than that given by FL theory (Eq. (19)). That 
 the measured drag is larger could be explained by the interaction-normalized 
 interwire backscattering probability, which should be larger than the bare 
 one (Eq. (34)). 

 Comparison of Figs. 9(a) and 9(b) show that the influence of the magnetic 
 field on the temperature dependence is not significant. This may signify 
 that Zeeman spin splitting at B $ \le $ 1 Tesla is not important yet, 
 otherwise the exponent $x$ should change (Eq. (35)). Indeed, a clear signature 
 of spin splitting in the measured conductance staircase was not observed at 
 this field. The magnetic field, however, increases
 $v_F$ for the same position of the Fermi level. This makes
 $L_{T}$\textit{} larger at the same temperature compared to that in  
 zero field. This can explain why in a magnetic field a deviation from the 
 power-law dependence occurs at a higher temperature (Fig.9(b)). 

 We have interpreted above the experimental data in terms of Coulomb drag 
 only. Considering the large interwire separation for which the drag 
 measurements were made, one can not rule out the possibility of an acoustic 
 phonon-mediated drag (PMD) contribution to the measured drag resistance. 
 Recent theoretical work \cite{raichev} \cite{muradov} on 1D PMD based
 on Fermi liquid description  
 predicts that PMD is negligible compared to Coulomb drag for 2$k_{F}d <5$. 
 Also, for a dual-wire sample shown in Fig.5, $R_{D}$ should increase 
 exponentially with temperature in the range 100 - 600 mK and does not 
 decrease exponentially with interwire separation $d$. The data shown in Figs.7 
 and 9 qualitatively contradict these predictions. This allows us to conclude 
 that the PMD contribution, if present at all, is insignificant.

\section{V. Future prospects} \label{sec-future}


 It is quite obvious from the content of Sec. IV that substantial 
 experimental work remains to be done to gain a comprehensive understanding 
 of the physics of Coulomb drag between interacting 1D electron systems and 
 to explore the conditions under which such systems behave as a Fermi liquid 
 or a Luttinger liquid. Since the measurement of the Coulomb drag also 
 provides a new experimental tool to probe the LL state that can't be done 
 from the measurement of the conductance alone, extensive experimental work 
 on the subject is needed to put the LL model of interacting 1D systems on a 
 firm footing. Though the theory of Coulomb drag has considerably outstripped 
 experimental work, many open questions need to be addressed in the 
 theoretical area as well.

 On the experimental side, work should be focused on measurements that can 
 distinguish between a LL and a FL state and can provide information about 
 the existence and the nature of the LL state. This is an extremely important 
 area for condensed matter physics. The few papers published so far, claiming 
 to have observed a Luttinger liquid, have not been convincing. In this 
 respect, it would be highly interesting to study the drag between spin 
 polarized systems, since the LL theory predicts different exponents for spin 
 polarized and unpolarized cases and manifestation of the spin effect should 
 be quite different in the Fermi liquid and the Luttinger liquid state. 
 Another interesting experimental possibility is to study drag when the wire 
 length $L$ falls below the thermal length $L_{T}$ to investigate if the drag 
 resistance $R_{D}$ decays $ \propto  \quad T^{2}$ as predicted by LL liquid 
 theory. When the number of electrons in the wires is very small (Sec. IV), 
 one should expect relatively large fluctuations of the drag current or 
 voltage, such as shot noise
 \cite{gurevich_muradov2000}\cite{trauzettel}, and possible reversal
 of the sign of drag leading to negative drag
 \cite{mortenson}. Observation of 
 this noise can also 
 provide valuable information on correlated electron state. One could also 
 envision a search for 1D spin Coulomb drag \cite{damico}. Finally, it is also 
 important to study acoustic phonon-mediated drag (PMD) \cite{raichev}
 \cite{muradov} since under  
 certain conditions it can be comparable to and even larger than the Coulomb 
 drag. If such a PMD is present in the experimental measurements, one has to 
 find ways to separate it from the Coulomb drag.

 The theory of Coulomb drag based on the LL model is far from mature and many 
 open questions need to be addressed such as the effect of disorder, the 
 influence of tunneling between the wires, etc.. Alhough the power-law 
 temperature dependence of the drag resistance is a signature of the 
 Luttinger liquid state, a careful analysis of various limiting cases based 
 on the Fermi liquid approach should be carried out to make sure that under 
 no circumstances it can give a similar temperature dependence. It is equally 
 important to investigate the physical situations and interactions (within 
 the wires and with the reservoirs) that favor transition of the Fermi liquid 
 into the Luttinger liquid and vice versa.

 \end{document}